# Synthesis of nanoparticles of the giant dielectric material, $CaCu_3Ti_4O_{12}$ from a precursor route.

P. Thomas,[a,b] K. Dwarakanath,[a] K.B.R. Varma[b] and T.R.N. Kutty.[b*]

[a] Dielectric Materials Division, Central Power Research Institute, Bangalore : 560 080, India
[b] Materials Research Centre, Indian Institute of Science, Bangalore: 560012, India,

**Abstract:**

A complex oxalate precursor, $CaCu_3(TiO)_4(C_2O_4)_8 \cdot 9H_2O$, (CCT-OX), was synthesized and the precipitate that obtained was confirmed to be monophasic by the wet chemical analyses, X-ray diffraction, FT-IR absorption and TG/DTA analyses. The thermal decomposition of this oxalate precursor led to the formation of phase-pure calcium copper titanate, $CaCu_3Ti_4O_{12}$, (CCTO) at $\geq 680^oC$. The bright-field TEM micrographs revealed that the size of the as synthesized crystallites to be in the 30-80nm range. The powders so obtained had excellent sinterability resulting in high density ceramics which exhibited giant dielectric constants upto 40,000 (1 kHz) at $25^oC$, accompanied by low dielectric losses, < 0.07.

## 1. INTRODUCTION

Electroceramics associated with giant dielectric constants are in increasing demand owing to the miniaturization of electronic devices. Oxides with the perovskite and related structures are well known for their high dielectric constants. The titanate compound $CaCu_3Ti_4O_{12}$ (CCTO) belongs to a family of multinary oxides of the type, $ACu_3Ti_4O_{12}$ (where A= Ca or Cd) and has been first reported in the year 1967 [1]. This composition has been extended with the general formula, $[AC_3] (B_4) O_{12}$, [where A= Ca,Cd,Sr,Na or Th; B = Ti or (Ti + $M^{5+}$), in which M= Ta, Sb or Nb; and C= $Cu^{2+}$ or $Mn^{3+}$] [2]. The crystal structure of $CaCu_3Ti_4O_{12}$ has been refined with the space group Im3, lattice parameter a ≈ 7.391 $\overset{o}{A}$, Z=2) and remains centrosymmetric body centered cube (bcc) over a wide range of temperatures [2]. CCTO has attracted considerable attention recently due to its unusually high dielectric constant ($\varepsilon \approx 10^4$-$10^5$) which is nearly independent of frequency (upto 10 MHz) and low thermal coefficient of permittivity (TCK) over a wider range of temperature (100-600K) [3,4]. CCTO was prepared by the solid-solid reactions between the stoichiometric starting materials of $CaCO_3$, $TiO_2$ and CuO. The mixtures are calcinated at high temperature for longer duration (typically 1000-1050°C for 24-48h) with repeated intermediate grindings [3,5,6]. This method of preparation is very cumbersome often requiring much longer duration of reaction, and temperature approaching the melting point of CuO. Besides, this procedure suffers from the disadvantages of chemical inhomogeneity leading to coarse particle size for the product. In contrast, the wet chemical syntheses routes offer homogeneous products at lower temperatures in shorter durations [7-9]. There a few research papers exist, indicating that CCTO could be prepared by routes other than the solid state reactions [10-14]. The powders prepared by the pyrolysis of the co-precipitated oxalates at >950°C for 10h yielded CCTO with $CaTiO_3$ + CuO as the impurity phases [14]. In order to avoid difficulties in obtaining phase-pure CCTO powders at relatively lower temperatures, we have developed a precursor oxalate route and found to be a very convenient method for achieving chemical homogeneity.

---

[*] Corresponding author: E-mail: <kutty@mrc.iisc.ernet.in> Tel. +91-80-2293-2784; Fax: +91-80-2360-7316.

## 2. EXPERIMENTAL
### 2.1 Preparation of the oxalate precursor complex

TiOCl$_2$ was obtained from the controlled reaction of ice-cold distilled water with TiCl$_4$ (titanium tetrachloride, 99.98%) (Merck, Germany). The other chemicals employed were calcium carbonate/calcium chloride (BDH; A.R.grade), cupric chloride/cupric nitrate (Fluka, proanalyse grade), oxalic acid (S.D. Fine Chemicals, analytical grade), ethanol or acetone (Nice, India; 99.5% pure). As the first step, titania gel [$TiO_2 \cdot xH_2O$ (92<x<118)] (0.4 moles) was prepared from the aqueous TiOCl$_2$(0.05M) by adding NH$_4$OH (aq) (at 25°C) till the pH reached ~ 8.0. NH$_4$Cl therein was washed off in the filter funnel. This gel was added to 0.4 or 0.8 moles of oxalic acid (2 M solution) (1:1 or 1:2 ratio of Ti: $C_2O_4^{2-}$) which was kept warm at ~40°C. The clear solution that obtained was reacted with CaCO$_3$ to yield titanyl oxalic acid + calcium titanyl oxalate in 3:1 ratio. It has been reported [15] that this aqueous solution remained clear without any precipitate formation. The mixed solution was cooled to 10°C to which cupric chloride (or cupric nitrate) dissolved in acetone+water (80/20) was added and stirred continuously. The thick precipitate of calcium copper titanyl oxlate (CCT-OX) was separated out by further addition of acetone. The precipitate was filtered, washed several times with acetone to make it chloride-free and dried in air. The precursor obtained by this method, remained finely divided with the minimum level of agglomeration.

## 2.2 Characterization Techniques

X-ray powder diffraction (XRD) studies were carried out with an XPERT-PRO Diffractometer (Philips, Netherlands) using Cu K$\alpha_1$ radiation ($\lambda$ = 0.154056 nm) in a wide range of 2$\theta$ (5$^o$ ≤ 2$\theta$ ≤ 85$^o$). Thermal analyses (DTA/TG) were done using the TA Instruments (UK), Model:SDTQ600, which recorded DTA and TG simultaneously with alumina as the reference material. The experiments were carried out at a heating rate of 10$^o$C min$^{-1}$ in flowing air atmosphere (flow rate:50cm$^3$ min$^{-1}$). Infrared spectra were recorded using a Perkin-Elmer FTIR spectrophotometer. Transmission Electron Microscopy (TEM) were carried out using FEI-Technai TEM (G-F30, Hillsboro, USA). The powder was cold-pressed into pellets of 12mm in diameter and 3mm in thickness using 3% poly vinyl alcohol (PVA) and 1% polyethylene glycol as binders and sintered at 1100$^o$C/2h. The densities of the sintered pellets were measured by the Archimedes principle using bromoform as the liquid medium. Scanning electron microscope (SEM) (Cambridge Stereoscan S-360) was employed to study the microstructure of the sintered pellets. The capacitance of the electroded (fired-on silver) pellets were measured as a function of frequency (100Hz–1MHz) using an impedance gain-phase analyzer (HP4194A) in the 100-600K temperature range. A two terminal capacitor configuration was employed for the measurement.

## 3. RESULTS AND DISCUSSION

### 3.1. Characterization of the complex oxalate precursor

The phase singularity of the complex oxalate precursor is confirmed by the wet chemical analyses, X-ray diffraction, and TG/DTA analyses. The wet chemical analyses of this air-dried precipitate (CCT-OX) gave: Ca: 2.98; Cu: 14.02; TiO: 19.06; C$_2$O$_4$: 52.08; H$_2$O: 11.87 % (by wt) : Cal. for CaCu$_3$(TiO)$_4$(C$_2$O$_4$)$_8$·9H$_2$O: Ca: 2.96; Cu: 13.99; TiO: 18.95; C$_2$O$_4$: 52.10; H$_2$O: 11.99 %.

### 3.1.1. X-ray powder diffraction analysis

Figure 1 shows the XRD pattern of the as-prepared precursor (CCT-OX) and for the residues obtained after thermal decomposition at selected temperatures. The XRD of the complex precursor (Fig. 1.a) has been compared with those of the oxalates of individual metal ions available from International Centre for Diffraction Data (ICDD). It is clear that the XRD of the complex precursor (CCT-OX) is quite different from those of the individual oxalates in the Bragg reflections and their relative intensities. The strong lines of copper oxalate hydrate (ICDD: 00-021-0297), calcium oxalate hydrate (ICDD: 00-003-0110), and titanium oxalate hydrate (ICDD: 00-032-1386) are different from those of the as-prepared complex precursor. This confirms the appearance of a new precursor phase as a result of the wet-chemical preparation.

### 3.1.2. Infrared spectrum of the complex oxalate precursor

The formation of single-phase oxalate is further confirmed by the FTIR analysis. The IR absorption spectrum of the precursor is quite different from those of the individual oxalates. There are some unique features for the spectrum presented in Figure 2(a) by way of multiplets in all the oxalate-related absorption bands. The broad absorption band in the 1750-1620 $cm^{-1}$ region can be resolved into multiple peaks of which the one around 1630 $cm^{-1}$ corresponds to the bending mode of $H_2O$(hydration). In contrast, the multiple absorption bands are better discernible in the region 1450-1200 $cm^{-1}$ which arise from the mixed vibrations (Figure 2.a). The fact that there are six or more clearly decipherable components herein indicates the presence of oxalates with both ($D_{2h}$) / ($D_{2d}$) and $C_{2V}/C_2$ point group symmetry, which is a direct evidence for the prevalence of both bridging (tetradentate) as well as the bidentate oxalate ions. The multiplets in the regions 950-750 $cm^{-1}$ and 550-380 $cm^{-1}$ arise from the in-plane bending modes of mixed vibrations as mentioned in the Figure 2(a) [16].

### 3.1.3. Thermal analyses

The thermal analyses (TG/DTA) were carried out on both the complex precursor (Figure 3) as well as on the individual oxalates (Figure 4a & 4b). The general features of TG/DTA curves obtained for the as-prepared complex precursor (Figure 3) are quite different from those of the individual oxalates( Figure 4.a & 4.b). A strong exothermic peak observed around 478°C and an endotherm at 753°C for calcium oxalate (Figure 4.b) are altogether absent for the complex precursor (Figure 3). Furthermore, the exothermic peak seen around 498°C for the titanyl oxalate (Figure 4.b) is absent in the case of the complex precursor (Figure 3). This implies that the thermal decomposition behaviour of the complex precursor is uniquely different and supports the fact that the precursor synthesized is not a mixture of individual oxalates but of single–phase calcium copper titanyl oxalate.

### 3.2. Thermal decomposition of the complex oxalate precursor

Thermal decomposition of this complex precursor is studied on similar lines for alkaline earth metal oxalates reported [15,17-20]. However the decomposition behavior is found to be complex in nature due to the back reaction of CO by reducing Cu(II) compounds to Cu(I) or Cu$^{(O)}$ which in turn converts to CuO in air. The TG of the as-prepared precipitate shows marked moss losses with increasing temperature in three steps. The total mass-loss observed from the TG is around 54.8%. The theoretically expected mass-loss is ≈ 54.6%, calculated from the initial composition $CaCu_3(TiO)_4(C_2O_4)_8 \bullet 9H_2O$, which is in good agreement with the experimentally obtained value (54.8%). The possible decomposition reactions are: (i) dehydration, (ii) decomposition of the oxalate to a complex oxycarbonate and (iii) decomposition of the intermediate carbonate to calcium copper titanate.

Dehydration takes place in two steps: In the temperature range 25-130$^o$C, eight moles of water of hydration are lost. The experimentally observed mass loss of 10.7% is in reasonable agreement with the calculated value of 10.6% for the removal of 8 moles of H$_2$O. Thermal decomposition of the oxalate takes place in 2-steps: The DTA (Figure 3) indicates only one major exothermic peak corresponding to the decomposition of oxalate. However, the careful scrutiny of the DTG curve (inset in Figure 3) reveals that the thermal decomposition takes places in two stages: The first exothermic reaction occurs in the temperature range of 180-200$^o$C. The residue at this stage has the composition: $CaCu_3(TiO)_4(C_2O_4)_5(CO_3)_3 \bullet CO_2$ as revealed by the chemical analyses of the residue after the isothermal heating at 185$^o$C for 24h and the mass-loss is ~ 6.1%. The gases evolved at this stage are carbon monoxide and H$_2$O (vap) as per the analyses using a gas chromatograph. The residue is grey to black which on dissolution in hydrochloric acid does not leave any insolubles by way of the carbon particles arising from the disproportionation of carbon monoxide. The second step of the oxalate decomposition is the main event occurring in the thermoanalytical experiments. This exothermic reaction occurs from 230-280$^o$C which involves simultaneous evolution of carbon monoxide and carbon dioxide resulting in the intermediate carbonate as the residue. The isothermal heating at 235$^o$C for 24 h yielded a residue with the composition: $CaCu_3Ti_4O_{11}(CO_3) \bullet CO_2$. The mass-loss of 37.5% as compared to 38% calculated for the formation of this residue. This stage of the oxalate decomposition involves a complex set of reactions which includes decomposition of the oxalate, oxidation of CO to CO$_2$ as also the possible disproportionation of CO to CO$_2$ + C. On further increasing the heat treatment temperature, the intermediate carbonate CaCu$_3$Ti$_4$O$_{11}$(CO$_3$)·CO$_2$ decomposes to the oxycarbonate with the release of carbon dioxide retained within the matrix. This gives rise to the residue with the composition of CaCu$_3$Ti$_4$O$_{11}$(CO$_3$). This occurs in the broad temperature range of 280 – 550$^o$C. The oxycarbonate $CaCu_3Ti_4O_{11}(CO_3)$ decomposes between 600-700°C with the evolution of carbon dioxide, giving rise to calcium copper titanate ($CaCu_3Ti_4O_{12}$). The final step in the DTG corresponds to this reaction. The observed mass-loss for the carbonate decomposition is 4.0% as against the calculated mass-loss of 4.2%. This difference in mass-loss may be attributed to the parallel reaction as per the decomposition scheme-II (presented later on in this section) that is taking place during the decomposition of the intermediate oxycarbonate; decomposition scheme-II also gives the same mass-loss as of the oxycarbonate.

The XRD pattern (Figure 1.b) for the residue $CaCu_3(TiO)_4(C_2O_4)_5(CO_3)_3 \bullet CO_2$ obtained from the isothermal heating at 185°C for 24 h. The intermediates that are formed depend upon the back reaction of the evolved gases which is influenced by the experimental conditions including the rate of heating and the accumulation of the evolved gases. In the case of isothermal treatment where the sample is slowly heated from room temperature to the set temperature, the X-ray pattern (Figure 1.b) does not show reflections corresponding to those of calcium carbonate, copper oxide or any one of the polymorphs of titanium dioxide (anatase/rutile/brookite). However, when the sample is introduced into the preheated furnace to >185°C, the exothermic decomposition of the oxalate sets in. The XRD reflections (Figure 1.c) of low intensity observed for CuO at this stage indicate that the bulk of the residue is amorphous to X-rays. The X-ray diffraction pattern (Figure 1.d) for the residue, $CaCu_3Ti_4O_{11}(CO_3) \bullet CO_2$, obtained after the second stage of oxalate decomposition between 230-280°C indicates reflections (low intensity) corresponding to CuO only, whereas the reflections corresponding to $TiO_2$ and $CaCO_3$ are not detected. This indicates that, at this stage, bulk of the residue is amorphous while the minor phase of CuO formed during the thermal decomposition is crystalline to x-rays. The X-ray diffraction pattern of the residue, $CaCu_3Ti_4O_{11}(CO_3)$, obtained at 550°C (Figure 1.e) shows the presence of only two phases: (1) CuO and (ii) $TiO_2$ (anatase) with low intensities. The line broadening observed for CuO is due to the nanometric size of the particles and intensities are very low compared to those observed (not shown) for 100 % CuO. The reflections corresponding to $CaCO_3$ are barely discernible. This again indicates that the major portion of the residue is not crystalline to X-rays. It is evident that during the thermal decomposition of the complex oxalate precursor leading to X-ray amorphous intermediate, a parallel reaction is taking place leading to the formation of $CaCO_3$+ 3CuO+ 4$TiO_2$. Furthermore, when the precursor oxalate is thermally decomposed at 550°C in shallow alumina trays, the XRD of the residue remained totally amorphous to X-rays. It is confirmed that the phase-pure calcium copper titanate (CCTO) is formed when the complex precursor is isothermally heated above 680°C (Figure 1.f). Figure 5 presents the bright field TEM image of the phase-pure CCTO powders obtained from the thermal decomposition of the oxalate precursor above 680°C. The TEM micrograph reveals that the some of the particles are well-faceted and the sizes of the crystallites are in the 30-80nm range.

The infrared (i.r) spectra (Figure 2.b) of the residue, $CaCu_3(TiO)_4(C_2O_4)_5(CO_3)_3 \cdot CO_2$, from the first stage of oxalate decomposition exhibit the absorptions of both oxalate and carbonate groups. The intensities as well as the multiplicity of the oxalate absorption bands have diminished as compared to those of the as-prepared precursor. In addition, a sharp band prevails around 2340 cm$^{-1}$ which can be assigned only to the asymmetric stretching mode of carbon dioxide retained in the matrix of the X-ray amorphous residue [17]. The i.r.spectrum (Figure 2.c) of $CaCu_3Ti_4O_{11}(CO_3) \bullet CO_2$ from the second stage of oxalate decomposition shows the prevalence of ionic carbonate ($\upsilon_{asy}$ at 1509 and 1401 cm-1) as also carbon dioxide ($\upsilon_{asy}$ at 2338 cm$^{-1}$) retained within the amorphous matrix. The i.r.spectrum of the residue, $CaCu_3Ti_4O_{11}(CO_3)$, from 550°C, (Figure 2.d) shows the absorption band around 1436 cm$^{-1}$ confirming the prevalence of oxycarbonate [17] at this stage. There is no absorption band in the region 2200- 2400 cm$^{-1}$, indicating the escape of entrapped carbon dioxide. The i.r.spectrum (Figure 2.e) of the phase-pure CCTO does not exhibit absorption bands around 1436 cm$^{-1}$ corresponding to the stretching mode of the carbonate, thereby confirming the complete decomposition of the oxycarbonate at this stage. Further, there are absorption bands in the region 380-700 cm$^{-1}$ arising from the mixed vibrations of $CuO_4$ and $TiO_6$ groups prevailing in the CCTO structure.

Based on the above observations, the following two schemes are proposed for the decomposition of CCT-OX:

**Decomposition Scheme I (major process):**

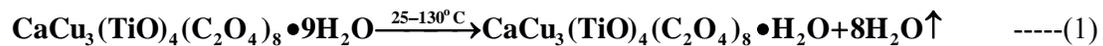
$$CaCu_3(TiO)_4(C_2O_4)_8 \bullet 9H_2O \xrightarrow{25-130°C} CaCu_3(TiO)_4(C_2O_4)_8 \bullet H_2O + 8H_2O \uparrow \quad \text{-----(1)}$$

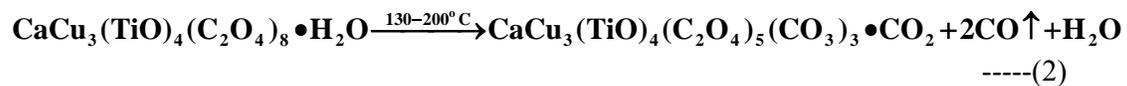
$$CaCu_3(TiO)_4(C_2O_4)_8 \bullet H_2O \xrightarrow{130-200°C} CaCu_3(TiO)_4(C_2O_4)_5(CO_3)_3 \bullet CO_2 + 2CO \uparrow + H_2O$$
$$\text{-----(2)}$$

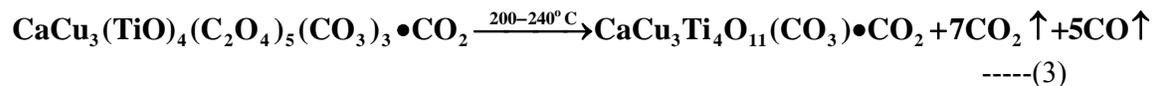
$$CaCu_3(TiO)_4(C_2O_4)_5(CO_3)_3 \bullet CO_2 \xrightarrow{200-240°C} CaCu_3Ti_4O_{11}(CO_3) \bullet CO_2 + 7CO_2 \uparrow + 5CO \uparrow$$
$$\text{-----(3)}$$

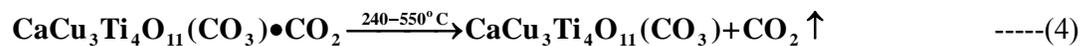
$$CaCu_3Ti_4O_{11}(CO_3) \bullet CO_2 \xrightarrow{240-550°C} CaCu_3Ti_4O_{11}(CO_3) + CO_2 \uparrow \quad \text{-----(4)}$$

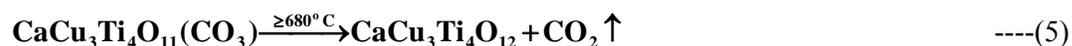
$$CaCu_3Ti_4O_{11}(CO_3) \xrightarrow{\geq 680°C} CaCu_3Ti_4O_{12} + CO_2 \uparrow \quad \text{----(5)}$$

**Decomposition Scheme II ( minor process ) :**

$$CaCu_3Ti_4O_{11}(CO_3) \bullet CO_2 \xrightarrow{240-550^oC} CaCO_3 + 3CuO + 4TiO_2 \qquad \text{-----(6)}$$

$$CaCO_3 + 3CuO + 4TiO_2 \xrightarrow{\geq 700^oC} CaCu_3Ti_4O_{12} + CO_2 \uparrow \qquad \text{-----(7)}$$

### 3.3 Dielectric characteristics of CCTO ceramics

The figure 6 shows the frequency (100Hz-100 kHz) dependence of room-temperature dielectric constants of CCTO ceramics sintered at 1100°C processed from the precursor derived powders. These disks have improved sinter density of >96% with the grain size varying from 40-100 $\mu$ m even though the duration of sintering has been shorter (~ 2h). This is indicative of the enhanced reactivity of the powders by way of high sinterability. The anomalously high dielectric constants of these ceramic samples are in the same range of those of the previous publications [3,4, 10-14]. At higher frequencies (>300kHz), the dielectric constant decreased drastically whereas the corresponding dielectric loss increased. Indeed, the dielectric constants of CCTO ceramics obtained by the present method are higher than those reported in the literature for CCTO ceramics [3,11,12]. The temperature dependence of dielectric constant measured at 1 kHz indicated that it is nearly constant in the temperature range of 100-600K which is consistent with that reported in the literature [3,4].

### 4. CONCLUSIONS

A wet chemical method has been developed for the preparation of complex oxalate precursor, $CaCu_3(TiO)_4(C_2O_4)_8 \bullet 9H_2O$. The precursor gives rise to phase-pure nanoparticles of $CaCu_3Ti_4O_{12}$ (CCTO) powders with a crystallite size varying from 30-80 nm when heat treated at > 680°C. The powders derived from the oxalate precursor have excellent sinterability resulting in high density ceramics which exhibited giant dielectric constants upto 40,000 (1 kHz) at 25°C, accompanied by low dielectric losses.

**Acknowledgements**

The management of Central Power Research Institute are acknowledged for the financial support (CPRI Project No.5.4.49).

# Figure Captions

**Figure. 1**. X-ray diffraction patterns of the (a) as prepared complex precursor; (b) slow heating at 185°C, (c) fast heating at 185°C, after heat treating at (d) 235°C, (e) 550°C, and (f) 700°C ( phase-pure CCTO); (g) CCTO from the ICDD data file card no. 01-075-1149.

**Figure. 2**. FTIR spectra of complex precursors (a) as prepared precursor, (b) at 185°C, (c) at 235°C , (d) at 550°C and (e) for phase-pure CCTO.

**Figure. 3**. Simultaneous thermal analysis ( DTA/TG ) for the complex precursor CCT-OX ( heating rate of 10°C min$^{-1}$ ) in air atmosphere (flow 50cm$^3$ min$^{-1}$). The inset shows the differential thermogram (DTG) of the complex precursor in the range of 220-325°C.

**Figure .4**. (a) TG curves of the individual oxalates namely, calcium oxalate, titanyl oxalate and copper oxalate, (b) DTA of the individual oxalates, calcium oxalate, titanyl oxalate and copper oxalate.

**Figure. 5.** Bright field TEM image of phase-pure CCTO obtained above 680°C with particle dimensions ranging from 30-80nm.

**Figure. 6**. Frequency dependence of the dielectric constant (25°C) and the loss factor (D) of the CCTO ceramics sintered at 1100°C (2h) from the powders obtained by the precursor route and compared with the reported values.

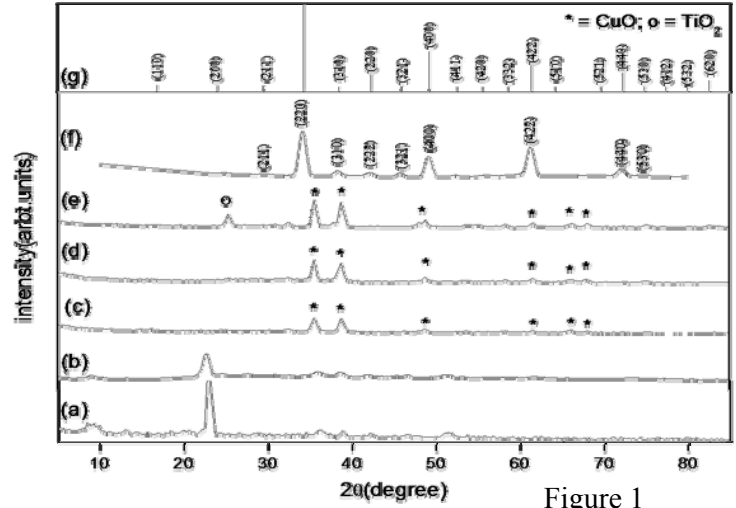

Figure 1

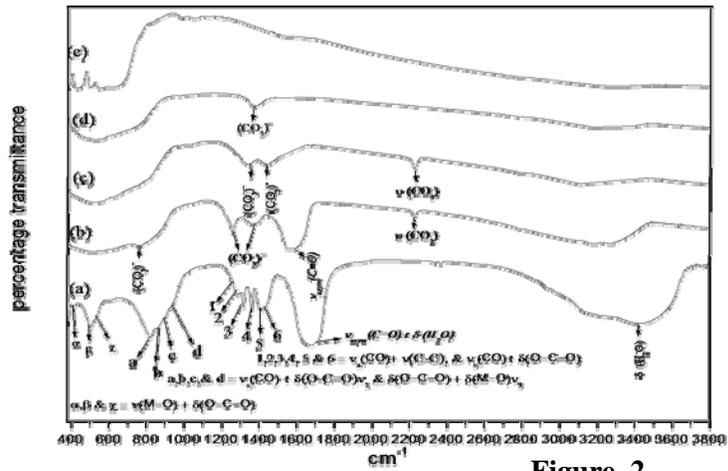

**Figure. 2**

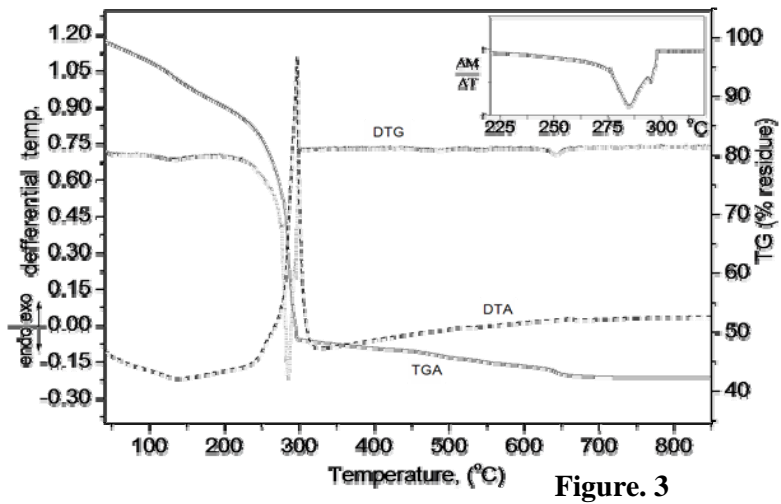

**Figure. 3**

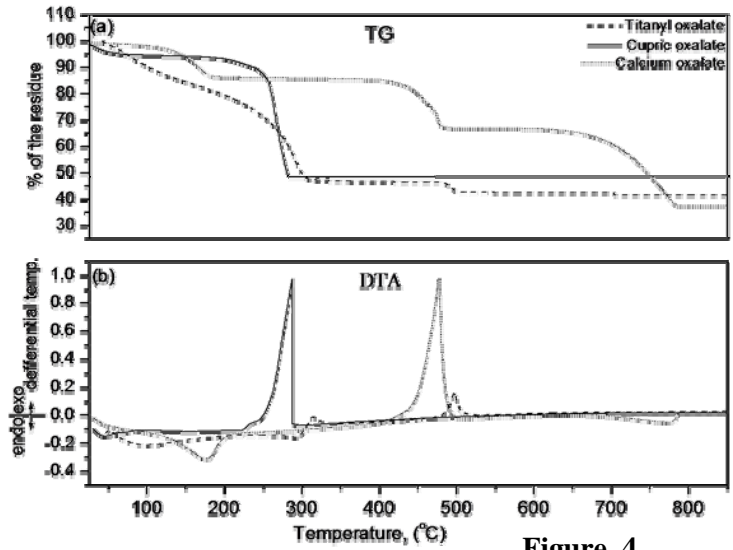

**Figure. 4**

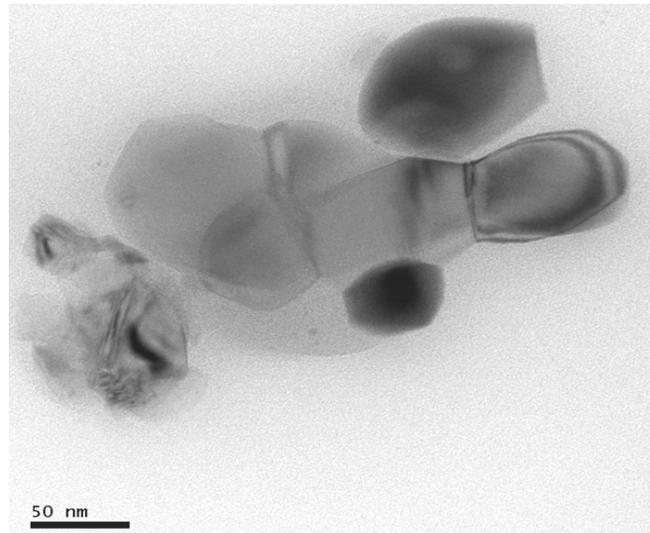

**Figure. 5**

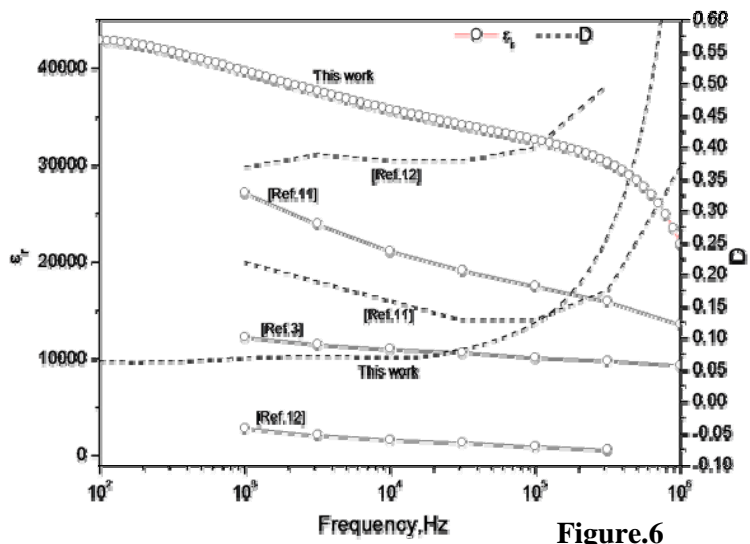

**Figure. 6**